\documentclass[12pt]{article}
\usepackage{amsmath,amssymb}
\usepackage[title]{appendix}
\usepackage{graphicx}
\graphicspath{ {./images/} }
\usepackage[font=small,labelfont=bf]{caption}
\usepackage{amsmath}
\usepackage{hyperref}
\usepackage{float}
\usepackage{pgfplots}
\pgfplotsset{compat=1.15}
\usepackage{mathrsfs}
\usepackage{pgf,tikz}
\usetikzlibrary{arrows}

\usepackage{color, soul}
\usepackage{xcolor}
\usepackage{caption}
\sethlcolor{yellow}
\newcommand {\be}{\begin{equation}}
 \newcommand {\ee}{\end{equation}}
 \newcommand {\bea}{\begin{array}}
 
 \newcommand {\eea}{\end{array}}

\textwidth=6.5in \hoffset=-.75in \voffset=-1in \numberwithin{equation}{section}
\numberwithin{figure}{section}

\def\0{{(0)}}
\def\1{{(1)}}
\def\2{{(2)}}

\def\<{\langle }
\def\>{\rangle }
\def\[{\left[}
\def\]{\right]}

\begin{document}
\begin{titlepage}

\vskip1cm
\begin{center}
{~\\[140pt]{ \LARGE {\textsc{Analytic Derivation of the minimal entanglement wedge cross section in the GMMG/GCFT flat holography    }}}\\[-20pt]}
\vskip2cm

\end{center}
\begin{center}
{Mohammad Reza Setare$^\dagger$ \footnote{E-mail: rezakord@ipm.ir}\hspace{1mm} ,
Meisam Koohgard \footnote{E-mail: m.koohgard@modares.ac.ir}\hspace{1.5mm} \\
{\small {\em  {Department of Science,\\
 Campus of Bijar, University of Kurdistan, Bijar, Iran }}}}\\
\end{center}
\begin{abstract}
We focus on a proper candidate for the entanglement wedge in asymptotically flat bulk geometries that are described by the generalized minimal massive gravity (GMMG) in the context of the flat holography. To this end, we describe the boundary by two dimensional Galilean conformal field theory (GCFT) at the bipartite mixed state of the two disjoint intervals. We derive an analytic expression for the minimal entanglement wedge cross section (EWCS) in the GMMG/GCFT framework. Our result provides an independent derivation that precisely matches previous computations of holographic entanglement negativity, thereby offering a powerful consistency check and validating both approaches within the GMMG/GCFT framework.
             \end{abstract}

\vspace{1cm}

Keywords: Entanglement wedge cross section, GMMG, Holographic entanglement negativity, bipartite mixed state.

\vspace{0.5cm}

\noindent\textbf{Comments:} This work is an extended version of arXiv:2203.08799 [hep-th] with significant updates to analysis and results.

\end{titlepage}

\tableofcontents

\section{Introduction}	
The entanglement entropy is a reliable measure to determine the quantum entanglement between the bipartite pure states. This measure is not a proper one for the mixed quantum states; the quantum correlations have more involved structures than the pure states \cite{Hors,Cal01}. A computable measure of the entanglement has been introduced in \cite{Vid01} (see also \cite{Vid02,Per,Eis,Pl01}) to characterize the correlation between the mixed states. This is a logarithmic negativity that can be computed by taking the trace norm of the mixed density matrix which is a partially transposed matrix related to the subsystems. A classical approach based on a replica method to compute the negativity in QFT has been introduced in \cite{Cal01}  and this method has been used for the various bipartite states in two dimensional conformal field theories ($CFT_2$'s) in \cite{Cal02,cft1,cft2,cft3,cft4}.

A holographic computation of the entanglement negativity \cite{hol01,hol02,hol03}  has been motivated by the holographic conjecture of Ryu and Takayanagi \cite{RT01,RT02} in computation of the universal part of the entanglement entropy of a $CFT_d$ which is dual to an $AdS_{d+1}$ in the bulk side of the duality. In the Ryu and Takayanagi's conjecture, a codimension-2 minimal surface (as RT surface) in the bulk is homologous to the subsystem in the boundary and its area is considered proportional to the universal part of the entanglement entropy. Along with the progress in the characterization of the holographic entanglement entropy in $CFT_d$'s, a proposal for the holographic entanglement negativity in the $AdS_{2+1}/CFT_2$ was developed in \cite{hol02,HEN01,HEN02}. This proposal was based on finding out some geodesics in the bulk homologous to some combinations of the intervals in the boundary.

Another approach to compute the holographic entanglement negativity have been introduced in \cite{Ku01} using the minimal cross sectional area of entanglement wedges. The entanglement wedge \cite{EW01,EW02,EW03}  is a natural region in the bulk that is distinct from the causal wedge \cite{EW01,EW02,EW03,CW01,CW02,CW03} which are both corresponding to the boundary configuration. The former is a bulk region that its dual in the boundary is the reduced density matrix of the CFT and the minimal entanglement wedge cross section has been proposed as a measure of the entanglement of purification (EoP) \cite{EW04,EW05,EW06,EW07}. In the theory of quantum information, the entanglement of purification that receives both classical and quantum contributions is a proper measure for both of the classical and quantum correlations in the bipartite mixed states \cite{EP01}. This feature of the EoP is in contrast to the entanglement entropy.  The minimal entanglement wedge cross section as a measure in relation to other entanglement measures, for example, the odd entanglement entropy \cite{OE} and the reflected entropy \cite{RE01,RE02,RE03,RE04} is an interesting quantity that could be investigated in the quantum entanglement.

The two dimensional Galilean conformal field theories ($GCFT_2$) were proposed as the non-relativistic version of the corresponding conformal field theories in two dimensional spacetime by using the Inonu-Wigner contraction of the $CFT_2$ algebra \cite{GC01,GC02,GC03,GC04}. Computation of the entanglement entropy of CFTs in \cite{GC04,GC05} was extended to find out the entanglement negativity of the $GCFT_2$ in \cite{GC06} through the replica technique. The non-relativistic flat holography between the bulk gravity and boundary
quantum field theory has been extended in \cite{holG01}. The three dimensional generalized minimal massive gravity (GMMG) \cite{GMMG01} could be an option for the bulk description of the two dimensional Galilean conformal algebra (2d GCA) with asymmetric central charges. GMMG model is a theory that avoids the bulk-boundary clash and, as a Minimal Massive Gravity (MMG) model \cite{MMG01}, it has some positive energy excitations around the vacuum that are the maximally $AdS_3$. The GMMG's central charges are positive in the dual CFT. In contrast to GMMG, the Topologically Massive Gravity (TMG) \cite{TMG01,TMG02} and the cosmological extension of New Massive Gravity \cite{NMG} that are constructed previously with local degrees of freedom
in three dimensional spacetime could not avoid the aforementioned class. GMMG model is a general extension of MMG that can be constructed from  pure Einstein gravity by adding the Chern-Simons (CS-) term and another term with a negative cosmological constant.

In the study of the proper measures for the entanglement between the mixed states, the holographic entanglement negativity of the bipartite states in $GCFT_2$ has been considered in \cite{Seng01}. The authors in \cite{Seng01} investigated the asymptotically flat (three dimensional) geometries in the bulk side of the holography and they considered the three dimensional Einstein gravity and the topologically massive gravity (TMG) dual to a class of $GCFT_2$'s. Their study on the entanglement negativity was based on the evaluation of some extremal curves homologous to the appropriate configurations of the bipartite state in the boundary of the holography. These are generalizations of the RT surfaces and are known as the swing surfaces \cite{sw01,sw02}. In \cite{GMMG02}, we extended the measurement of the entanglement negativity to another flat and non-relativistic holography that we considered the GMMG in the asymptotically flat background of the bulk and the $GCFT_2$ in the bipartite configuration of the boundary. We utilized the replica technique in the computation of the entanglement negativity and the results match with the results in \cite{Seng02} in the large central charge limit.

The computation of the entanglement negativity in the $AdS_3/CFT_2$ duality has been developed through the computation of the bulk entanglement wedge cross section (EWCS) in \cite{Ku01,Ku02}. The development of this approach to flat holography has been done by a novel construction in \cite{Seng03}. In this construction, a new approach to computation of the EWCS has been extended in some asymptotically flat three dimensional gravities in the bulk side of the duality. Based on the structure  of the minimal entanglement wedge cross section in the $TMG/GCFT_2$ framework \cite{Seng03}, we analytically derive its counterpart in the $GMMG/GCFT_2$ and we find the holographic entanglement negativity in $GMMG/GCFT_2$ framework. In particular, we present an analytic derivation of the minimal entanglement wedge cross section in the $GMMG/GCFT_2$ holographic framework, based on the action of a conical defect worldline in the bulk geometry .As a consistency check we find the result in this work the same as the result on the $GMMG/GCFT_2$ on our previous work in \cite{GMMG02}. This result was previously obtained via replica field theory in \cite{GMMG02}, but here we derive it independently from the bulk gravitational action. 

In section \ref{sec:2} we study the minimal EWCS in the AdS/CFT scenario. In the following, we introduce the GMMG model as the model describing the three dimensional asymptotically flat space geometry of the bulk side of the GMMG/GCFT duality and we present an argument on the structure of the minimal EWCS of the GMMG and its relation with the result in the TMG case. In the section \ref{sec:3}, we focus on the minimal EWCS in the cases where bipartite mixed with two disjoint intervals are considered at the boundary. First, we find the minimal EWCS at the Minkowski space of the bulk that is dual to the mixed bipartite state of $GCFT_2$ in the vacuum. Then the minimal EWCS at the flat space geometry of the bulk is introduced in the following of this section. In the latter case, the boundary will be introduced by two disjoint mixed intervals in the $GCFT_2$ at a finite temperature state. In the last section, we present  our results of this study.

\section{EWCS in asymptotically flat space GMMG}\label{sec:2}

In this section, we aim to derive a bulk gravitational expression for the minimal entanglement wedge cross section (EWCS) in the context of flat space holography based on Generalized Minimal Massive Gravity (GMMG). We begin by reviewing the known relation between EWCS and holographic negativity in AdS/CFT, and then proceed to formulate the corresponding construction in the $GMMG/GCFT_2$ framework. Our goal is to derive an explicit formula, presented in (\ref{TMG-E01}), from the worldline action of a conical defect propagating in the bulk geometry.

\subsection*{2.1 Minimal EWCS in AdS/CFT}
\begin{figure}[!h]
\centering
\captionsetup{width=.8\linewidth}
\definecolor{ffqqqq}{rgb}{1,0,0}
\definecolor{qqqqff}{rgb}{0,0,1}
\begin{tikzpicture}[line cap=round,line join=round,>=triangle 45,x=1cm,y=1cm]
\clip(-3,-0.7) rectangle (3,2.2);
\draw [line width=1.2pt,color=qqqqff] (-2,0)-- (-1,0);
\draw [line width=1.2pt] (-1,0)-- (1,0);
\draw [line width=1.2pt,color=qqqqff] (1,0)-- (2,0);
\draw [shift={(0,0)},line width=1.2pt,color=qqqqff]  plot[domain=0:3.141592653589793,variable=\t]({1*1*cos(\t r)+0*1*sin(\t r)},{0*1*cos(\t r)+1*1*sin(\t r)});
\draw [shift={(0,0)},line width=1.2pt,color=qqqqff]  plot[domain=0:3.141592653589793,variable=\t]({1*2*cos(\t r)+0*2*sin(\t r)},{0*2*cos(\t r)+1*2*sin(\t r)});
\draw [line width=1.2pt] (-2,0)-- (-3,0);
\draw [line width=1.2pt] (2,0)-- (3,0);
\draw [line width=1.2pt,dash pattern=on 1pt off 1pt,color=ffqqqq] (0,2)-- (0,1);
\draw (1.2834039262343875,0.030495181439616773) node[anchor=north west] {B};
\draw (-1.6481844140392636,-0.0742044021415849) node[anchor=north west] {A};
\draw (-0.014870910172515057,1.8522679357525258) node[anchor=north west] {$\Sigma_W^{min}$};
\end{tikzpicture}

\caption{The region bounded blue solid line is the entanglement wedge $M_{AB}$ for the bipartite mixed state on $A\cup B$ in the static $AdS_3$. The minimal entanglement wedge cross section is depicted as the red colored dashed line. }
\label{figEW001}
\end{figure}
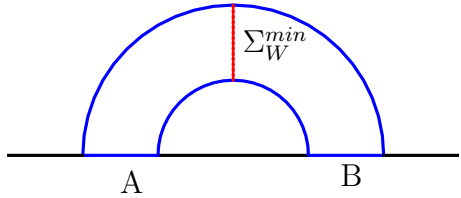

The relation between the minimal EWCS and the entanglement negativity in the $AdS_{d+1}/CFT_d$ duality plays an important role in our study in extending the relation to the GMMG/GCFT duality.
For this reason, in this section, we will briefly review some results in the EWCS in $AdS_{d+1}/CFT_d$ framework. Since our focus in this paper is on the bipartite mixed state of the boundary, we consider two disjoint intervals $A$ and $B$ in a $CFT_d$ at a slice with constant time. The static geometry of the bulk is considered with $AdS_{d+1}$. If $\Gamma_{AB}$ is the RT surface for the interval $A\cup B$, the entanglement wedge $M_{AB}$ is a region bounded blue solid line by the following boundary \cite{EW04} (Fig.\ref{figEW001})
\begin{equation}\label{EW.b}
  \partial M_{AB}=A\cup B\cup \Gamma_{AB}.
\end{equation}

The minimal EWCS can be defined as follows \cite{EW04}
\begin{equation}\label{EW01}
  E_W(A:B)=\frac{Area(\Sigma_{AB}^{min})}{4G_N}
\end{equation}
where $\Sigma_{AB}^{min}$ is a minimal surface that can split two entanglement wedges belonging to $A$ and $B$. The $G_N$ is Newton's constant. The minimal EWCS can be seen in Fig.\ref{figEW001} by the red colored dashed line area.

When the subsystems in the $CFT_d$'s have a spherical entangling surface, the holographic entanglement negativity $\mathcal{E}$ has the following direct relation with the minimal EWCS \cite{Ku01,Ku02}
\begin{equation}\label{HEN01}
  \mathcal{E}=\zeta_d E_W,
\end{equation}
where $\zeta_d$ is a constant that its relation for a vacuum state in a $CFT_d$ can be determined as follows \cite{cft3,Ren01}
\begin{eqnarray}\label{Xd01}
  \zeta_d &=& \bigg(\frac{1}{2}x_d^{d-2}(1+x_d^2)-1 \bigg), \nonumber \\
  x_d &=& \frac{2}{d}\bigg(1+\sqrt{1-\frac{d}{2}+\frac{d^2}{4}}  \bigg).
\end{eqnarray}

For the case of the $AdS_3/CFT_2$, the relation (\ref{HEN01}) is as follows
\begin{equation}\label{HEN02}
  \mathcal{E}=\frac{3}{2} E_W,
\end{equation}
where can be obtained by substituting $d=2$ into (\ref{Xd01}). In this case all the entangling surfaces of subsystems have the spherical geometries.

\subsection*{2.2 Worldline Action and Cone Geometry in GMMG}

The story in flat holography is somewhat different. The authors in \cite{Seng03}, given the relation between the minimal EWCS and the four-point functions of the twist fields inserted in the two disjoint intervals, claim that if the computational channels in the monodromy analysis \cite{Ku01} changes, a non-zero result for the EWCS can be obtained. The four-point functions are expanded based on the conformal blocks, and as the channel changes, the dominant term of the expansion as the channel changes. By changing the dominant term in the four-point function, a transition from zero to non-zero is obtained in the minimal EWCS. Such a transition has already been observed in the AdS/CFT case \cite{RE02,chan01,chan02}.  In \cite{Seng03} , we do not only see the non-zero minimal EWCS for the flat holography, but also the relation that is obtained for the EWCS in the holography has a relation with the entanglement negativity that was obtained before in \cite{Seng02}. The relationship between the minimal EWCS $E_W$ and the holographic entanglement negativity $\mathcal{E}$ is obtained in the cases of flat Einstein gravity and the TMG which both are dual to the $GCFT_2$. In both cases, the results are the same as the result in the $AdS_3/CFT_2$ framework.

In our previous work \cite{GMMG02} in the GMMG/GCFT framework, we obtained the holographic entanglement negativity with the help of the replica technique. In this paper, we present another approach for computation of the entanglement negativity with the help of the minimal EWCS. The GMMG model and the TMG model are similar in some respects in that they include the Chern-Simons and the Einstein gravity parts, but GMMG also includes the higher derivative gravity terms. The Lagrangian 3-form of the GMMG model can be written as follows \cite{GMMG01}
\begin{equation}\label{GM01}
  L_{GMMG}=L_{TMG}-\frac{1}{m^2}\big(f.R+\frac{1}{2}e.f\times f\big)+\frac{\alpha}{2}e.h\times h
\end{equation}
 where $L_{TMG}$ is the Lagrangian of the TMG and $R$ is the Ricci scalar. $f$ is an auxiliary one-form field, and $m$ is a mass parameter term. $e$  is a dreibein and $h$ is the auxiliary field. The Chern-Simons (CS-) term in the $L_{TMG}$ has the following form in the first order formalism
 \begin{equation}\label{CS01}
   CS-term: \frac{1}{2\mu}(\omega.d\omega+\frac{1}{3}\omega.\omega\times\omega)
 \end{equation}
where $\omega$ is dualised spin-connection. The algebra of the asymptotic conserved charges of asymptotically $AdS_3$ spacetimes in the context of GMMG is isomorphic to two copies of the Virasoro algebra with the following right and left central charges \cite{GMM03}
\begin{equation}\label{Vir.cen2}
  c_{GMMG}^+=\frac{3l}{2G}(-\sigma-\frac{\alpha H}{\mu}-\frac{F}{m^2}+\frac{1}{\mu l}),~~~c_{GMMG}^-=\frac{3l}{2G}(-\sigma-\frac{\alpha H}{\mu}-\frac{F}{m^2}-\frac{1}{\mu l}),
\end{equation}
where $H$ and $F$ are some constants. By taking In\"on\"u-Wigner contraction \cite{GC01,GC02}, we can find the asymptotic symmetry group as a
 GCA with the following central charges
 \begin{eqnarray}\label{cLM-GMMG}
  c_L &=& c_{GMMG}^+-c_{GMMG}^-=\frac{3}{\mu G}, \nonumber \\
  c_M &=& \frac{(c_{GMMG}^++c_{GMMG}^-)}{l}=\frac{3}{G}(-\sigma-\frac{\alpha H}{\mu}-\frac{F}{m^2}).
\end{eqnarray}

Based on the argument presented in the appendix \ref{app}, we are led to make a claim that the minimal EWCS of the GMMG in the context of the asymptotically flat geometry can have a similar structure with the minimal EWCS in the flat TMG. To provide an argument in support of this claim, we present an argument similar to the one we have already given in relation to negativity in \cite{GMMG02}. To this end, we follow the approach taken by the authors in \cite{sp02,Mald}. The authors
consider the bulk action with a conical singularity of $2\pi\epsilon $ along a worldline $C$ in
the bulk where $\epsilon$ is equal $n-1$. $n$ defines the number of the copies of the plane that are
sewn together to form the boundary interval in the holography. The entanglement wedge cross section can be related to the worldline action. The worldline metric could have the following form
\begin{eqnarray}\label{mCo}
  ds^2 &=& e^{\epsilon\phi(\gamma)}\delta_{ab}d\gamma^a d\gamma^b+(g_{zz}+K_a\gamma^a+...)dz^2 \nonumber\\
  && + e^{\epsilon\phi(\gamma)}U_a(\gamma,z)d\gamma^a dz,
\end{eqnarray}
where $z$ is a spacelike direction along the cone worldline and $\gamma^a$ are two perpendicular flat coordinates. $K_a$ as extrinsic curvatures are the expansion coefficients of $g_{zz}$ along $\gamma^a$ directions. $U_a$ are some arbitrary functions and $\phi(\gamma)$ is a function that is needed to regularize the cone. By the technique extended in \cite{sp02,Mald,Dong,Camp} , we find the regularized cone action in the GMMG as follows \cite{GMMG02}
\begin{eqnarray}\label{SCO01}
  S_{cone}|_{GMMG} &=& -\frac{\epsilon}{4G} (-\sigma-\frac{\alpha H}{\mu}-\frac{F}{m^2}) \int_{C}dz\sqrt{g_{zz}}\nonumber\\
  && -\frac{i\epsilon}{16\mu G}\int_{C}dz\epsilon^{ab}\partial_a U_b+\mathrm{O}(\epsilon^2)
\end{eqnarray}
where the action have been computed to the first order in $\epsilon$ in the expansion. The second term in (\ref{SCO01}) is related to the CS-term in the GMMG Lagrangian. By changing the first term of (\ref{SCO01}) in a covariant manner, the cone action can written as follows
\begin{eqnarray}\label{SCO02}
  S_{cone}|_{GMMG} &=& -\frac{\epsilon}{4G} (-\sigma-\frac{\alpha H}{\mu}-\frac{F}{m^2}) \int_{C}d\tau \sqrt{g_{\mu\nu}\big(X(\tau)\big)\dot{X}^{\mu}\dot{X}^{\nu}   }\nonumber\\
  && -\frac{i\epsilon}{16\mu G}\int_{C}dz\epsilon^{ab}\partial_a U_b+\mathrm{O}(\epsilon^2)
\end{eqnarray}
where $X^{\mu}(\tau)$ parameterizes the cone tip and $\dot{X}^{\mu}\equiv \frac{\partial X^{\mu}}{\partial\tau}$.

By the following normal vectors
\begin{eqnarray}\label{nvs}
  n_{1} &\equiv& \frac{\partial}{\partial\gamma^{1}},  \nonumber\\
  n_{2} &\equiv & \frac{\partial}{\partial\gamma^{2}},
\end{eqnarray}
the $CS$-term in the $S_{cone}$ can be written as follows
\begin{eqnarray}\label{SCO03}
  S_{cone}|_{GMMG} &=& -\frac{\epsilon}{4G} (-\sigma-\frac{\alpha H}{\mu}-\frac{F}{m^2}) \int_{C}d\tau \sqrt{g_{\mu\nu}\big(X(\tau) \big)\dot{X}^{\mu}\dot{X}^{\nu}  }\nonumber\\
  && -\frac{i\epsilon}{4\mu G}\int_{C}d\tau n_{2}.\nabla n_{1}+\mathrm{O}(\epsilon^2),
\end{eqnarray}
where the "$\nabla n$" is the covariant derivative along the worldline. By changing the normal vectors to the Lorentzian signature as follows
\begin{eqnarray}\label{EuLo}
  \tilde{n}_{1} &\equiv& i n_{1}=\partial_t, \\
  \tilde{n}_{2} &\equiv& n_{2},
\end{eqnarray}
the action could have the following form
\begin{equation}\label{SCO04}
  S_{cone}|_{GMMG} = -\epsilon\frac{c_M}{12} L^{extr} -
   \epsilon\frac{c_L}{12}\Delta\eta^{extr},
\end{equation}
where we have used the following relations for the extremal curve $L^{extr}$ and the extremized boost $\Delta\eta^{extr}$, respectively as follows
\begin{equation}\label{EC01}
  L^{extr}= \int_{C}d\tau \sqrt{g_{\mu\nu}\big(X(\tau)\big)\dot{X}^{\mu}\dot{X}^{\nu}   }
\end{equation}
and
\begin{equation}\label{EB01}
  \Delta\eta^{extr}=\int_{C}d\tau \tilde{n}_{2}.\nabla \tilde{n}_{1}+\mathrm{O}(\epsilon^2),
\end{equation}
$L^{extr}$ is the lengths of the extremal curve that is homologous to the entangling surfaces at the boundary. $\Delta\eta^{extr}$ is the extremized boost will be explained in the next section.

Given the effect that worldline action receives from the Chern-Simmons term in (\ref{SCO04}), we find that such an effect naturally leads to the following analytic expression for the minimal entanglement wedge cross section in flat GMMG

\begin{equation}\label{TMG-E01}
  E_W=\frac{c_M}{12}L^{extr}+\frac{c_L}{12}\Delta\eta^{extr}.
\end{equation}

This result is similar to the TMG/GCFT one in \cite{Seng03} and the effect of the differences of the GMMG model is reflected only in the central charges in the minimal EWCS relation and that there is no change in the computations of $L^{extr}$ and $\Delta\eta^{extr}$. To test the validity of our analytic result, we consider configurations with two disjoint intervals in the following section.
We find that by using the relation (\ref{TMG-E01}), the results are consistent with our previous work in \cite{GMMG02}.

\subsection*{2.3 Derivation of Equation (2.20) from the Cone Action}

In the previous subsection, we computed the worldline action of a conical defect propagating in the bulk of GMMG. Here, we provide a derivation of the relation (2.20), which expresses the minimal entanglement wedge cross section (EWCS) in terms of geometric data along the defect worldline.

In the context of holographic replica methods, it has been established that the entanglement entropy can be extracted from the gravitational action evaluated on a bulk geometry with a conical singularity of opening angle $2\pi(1-\epsilon)$. This idea was originally proposed by Lewkowycz and Maldacena~\cite{Mald} and later extended by Dong~\cite{Dong} and Camps~\cite{Camp} to higher-derivative gravity theories. These results imply that the leading $\mathcal{O}(\epsilon)$ contribution to the gravitational action encodes geometric information related to entanglement.

Following similar reasoning, the minimal entanglement wedge cross section, which is conjectured to be dual to entanglement of purification or holographic negativity, can be derived from the first-order variation of the gravitational action $S_{\text{cone}}$ with respect to $\epsilon$, i.e.,
\begin{equation}
	E_W = \left. \partial_\epsilon S_{\text{cone}} \right|_{\epsilon \to 0}.
\end{equation}
This method was applied in the context of Chern-Simons gravity and holographic negativity in AdS and flat holography in~\cite{Ku01, Seng03, sp01, sp02}.

In our case, using the cone action derived in Eq.~(2.14), the GMMG worldline action has the form
\begin{equation}
	S_{\text{cone}} = -\epsilon \left( \frac{1}{4G} \mathcal{A} L_{\text{extr}} + \frac{1}{4\mu G} \Delta \eta_{\text{extr}} \right) + \mathcal{O}(\epsilon^2),
\end{equation}
where $L_{\text{extr}}$ is the length of the extremal curve connecting the entanglement wedge boundaries, $\Delta \eta_{\text{extr}}$ is the extremized boost angle, and $\mathcal{A} = -\sigma - \frac{\alpha H}{\mu} - \frac{F}{m^2}$. This leads directly to
\begin{equation}
	E_W = \frac{1}{4G} \mathcal{A} L_{\text{extr}} + \frac{1}{4\mu G} \Delta \eta_{\text{extr}}.
\end{equation}

By using the expressions for the GMMG central charges (Eq.~2.9),
\begin{equation}
	c_M = \frac{3}{G} \mathcal{A}, \quad c_L = \frac{3}{\mu G},
\end{equation}
we arrive at the final expression:
\begin{equation}
	E_W = \frac{c_M}{12} L_{\text{extr}} + \frac{c_L}{12} \Delta \eta_{\text{extr}},
\end{equation}
which proves Eq.~(2.20) and confirms that it is not merely a conjecture, but a direct consequence of the conical singularity analysis in the GMMG framework.

\section{EWCS for two disjoint intervals}\label{sec:3}

The authors in \cite{Seng03} consider a general case for a bipartite mixed disjoint intervals in the $CFT_2$ which is dual to $AdS_3$. Through the following contraction on the spacetime coordinates at the boundary,
\begin{equation}\label{cont01}
  t\to t,~~~~x\to\epsilon x~~~(with~~~ \epsilon\to 0),
\end{equation}
they obtain a result of the minimal EWCS for the bulk could have an asymptotically flat geometry. Using the above limit, the GCA algebra generators can be built from the CFT Virasoro generators \cite{GC08}. The results of \cite{Seng03} on minimal EWCS of a generic form of two disjoint intervals can be compared with the results of \cite{Seng01} on the entanglement negativity of the intervals. By this comparison, it can be confirmed that when the geometry of the bulk is considered to be an asymptotically flat one, and there is a $GCFT_2$ at the boundary, the following relation can be used between the entanglement negativity and the minimal EWCS \cite{Seng03}
\begin{equation}\label{ENEW01}
  \mathcal{E}=\frac{3}{2}E_W.
\end{equation}

Since we have now derived an exlicit formula for the EWCS in the $GMMG/GCFT_2$ setup, we apply this result to the cofiguration of two disjoint intervals to extract the entanglement negativity.

\subsection{EWCS for two disjoint intervals in the vacuum}\label{sec:3.1}
To find the EWCS for the bipartite mixed states, we describe the boundary in the $GCFT_2$ vacuum that is dual to the bulk flat 3-dimensional spacetime. The bulk spacetime can be described by the Minkowski metric in Eddington-Finkelstein coordinates as follows
\begin{equation}\label{Mink01}
  ds^2=-du^2+dr^2+r^2d\phi^2,
\end{equation}
where $\phi$ is the angular coordinate and $u=t-r$ is the retarded time. The two intervals $A=\big[\big(u_1,\phi_1\big),\big(u_2,\phi_2\big) \big]$ and $B=\big[\big(u_3,\phi_3\big),\big(u_4,\phi_4\big) \big]$ are considered in the cylindrical coordinates. The planar coordinates $(x,t)$ in the boundary have the following relation with the cylindrical coordinates
\begin{equation}\label{pl.cyl01}
  x=e^{i\phi},~~~t=iue^{i\phi}.
\end{equation}

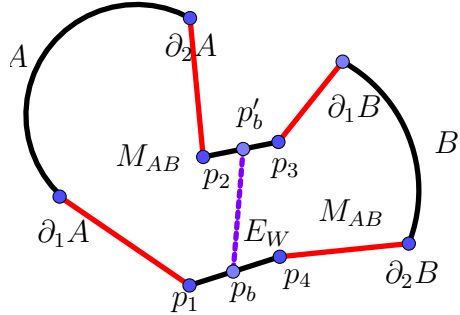
\begin{figure}[!h]
\centering
\captionsetup{width=.8\linewidth}
\definecolor{xfqqff}{rgb}{0.4980392156862745,0,1}
\definecolor{ffqqqq}{rgb}{1,0,0}
\definecolor{xdxdff}{rgb}{0.49019607843137253,0.49019607843137253,1}
\definecolor{ududff}{rgb}{0.30196078431372547,0.30196078431372547,1}
\begin{tikzpicture}[line cap=round,line join=round,>=triangle 45,x=1cm,y=1cm]
\clip(-3.2,-1) rectangle (3,4);
\draw [shift={(0.2818181818181815,1.19)},line width=2pt]  plot[domain=-0.37185607384858166:1.0743735733900155,variable=\t]({1*1.951562205342765*cos(\t r)+0*1.951562205342765*sin(\t r)},{0*1.951562205342765*cos(\t r)+1*1.951562205342765*sin(\t r)});
\draw [shift={(-1.5181818181818179,2.19)},line width=2pt]  plot[domain=1.0684939486969438:3.9551965009733876,variable=\t]({1*1.4728395018966276*cos(\t r)+0*1.4728395018966276*sin(\t r)},{0*1.4728395018966276*cos(\t r)+1*1.4728395018966276*sin(\t r)});
\draw [line width=2pt,color=ffqqqq] (-2.542975206611571,1.1057024793388348)-- (-0.8157024793388428,-0.07611570247934825);
\draw [line width=2pt,color=ffqqqq] (-0.8090909090909091,3.4809090909090905)-- (-0.6338842975206608,1.6329752066115628);
\draw [line width=2pt,color=ffqqqq] (1.2199840703542635,2.901268478909613)-- (0.36611570247933994,1.832975206611563);
\draw [line width=2pt,color=ffqqqq] (2.1,0.4809090909090906)-- (0.3842975206611581,0.30570247933883365);
\draw [line width=2pt] (-0.8157024793388428,-0.07611570247934825)-- (0.3842975206611581,0.30570247933883365);
\draw [line width=2pt] (-0.6338842975206608,1.6329752066115628)-- (0.36611570247933994,1.832975206611563);
\draw [line width=2pt,dotted,color=xfqqff] (-0.10241576605212899,1.7392689129052692)-- (-0.23558990898236942,0.10846556990680227);
\draw (-1.9429752066115709,1.9420661157024721) node[anchor=north west] {$M_{AB}$};
\draw (0.7479338842975221,1.2147933884297442) node[anchor=north west] {$M_{AB}$};
\draw (-1.1975206611570248,-0.02157024793389364) node[anchor=north west] {$p_1$};
\draw (-2.9793388429752077,0.942066115702471) node[anchor=north west] {$\partial_1A$};
\draw (-1.2884297520661159,3.4147933884297466) node[anchor=north west] {$\partial_2A$};
\draw (0.8752066115702494,2.5966115702479273) node[anchor=north west] {$\partial_1B$};
\draw (1.6388429752066136,0.396611570247925) node[anchor=north west] {$\partial_2B$};
\draw (0.12975206611570342,1.7602479338842902) node[anchor=north west] {$p_3$};
\draw (0.2933884297520672,0.2875206611570158) node[anchor=north west] {$p_4$};
\draw (-0.4157024793388424,0.10570247933883377) node[anchor=north west] {$p_b$};
\draw (-0.3247933884297515,2.560247933884291) node[anchor=north west] {$p'_b$};
\draw (-0.25206611570247867,0.942066115702471) node[anchor=north west] {$E_{W}$};
\draw (2.293388429752069,2.123884297520654) node[anchor=north west] {$B$};
\draw (-3.4157024793388446,3.269338842975201) node[anchor=north west] {$A$};
\draw (-0.7793388429752063,1.6147933884297445) node[anchor=north west] {$p_2$};
\begin{scriptsize}
\draw [fill=ududff] (2.1,0.4809090909090906) circle (2.5pt);
\draw [fill=ududff] (-0.8090909090909091,3.4809090909090905) circle (2.5pt);
\draw [fill=ududff] (-2.542975206611571,1.1057024793388348) circle (2.5pt);
\draw [fill=xdxdff] (1.2199840703542635,2.901268478909613) circle (2.5pt);
\draw [fill=ududff] (-0.8157024793388428,-0.07611570247934825) circle (2.5pt);
\draw [fill=ududff] (-0.6338842975206608,1.6329752066115628) circle (2.5pt);
\draw [fill=ududff] (0.36611570247933994,1.832975206611563) circle (2.5pt);
\draw [fill=ududff] (0.3842975206611581,0.30570247933883365) circle (2.5pt);
\draw [fill=xdxdff] (-0.10241576605212899,1.7392689129052692) circle (2.5pt);
\draw [fill=xdxdff] (-0.23558990898236942,0.10846556990680227) circle (2.5pt);
\end{scriptsize}
\end{tikzpicture}

\caption{The connected entanglement wedge construction for two disjoint intervals $A$ and $B$ in a $GCFT_2$ which is dual to an asymptotically flat spacetime that can be described by the Einstein gravity.}
\label{figEW02}
\end{figure}

If the asymptotically flat geometry of the bulk described by the pure Einstein gravity that only has one central charge $c_M$ in the dual field theory, the minimal EWCS can be obtained as follows
\begin{equation}\label{el01}
  E_{W1}=\frac{c_M}{12}L^{extr}(p_b,p'_b)
\end{equation}
where $L^{extr}(p_b,p'_b)$ is the length of the extremal curve which is homologous to A and B intervals (this extremal curve has been depicted by the blue dashed line in Fig.\ref{figEW02}). The non-trivial entanglement wedge has been shown by the $M_{AB}$ region in Fig.\ref{figEW02}. The red lines in Fig.\ref{figEW02} lie on the null planes extending from the endpoints of the intervals. $p_1$, $p_2$, $p_3$ and $p_4$ lie on the intersection of the null planes. The extremal curves connecting $p_1(p_2)$ and $p_4(p_3)$ lie on the null planes extending from $\partial_1A(\partial_2A)$ and $\partial_2B(\partial_1B)$. The curve that connects $p_1$ to $p_4$ is called $\gamma_{14}$ (and so is $\gamma_{23}$ for the points $p_2$ and $p_3$). $L^{extr}(p_b,p'_b)$ is the extremized length of the curve connects two arbitrary points $p_b$ and $p'_b$ on the $\gamma_{14}$ and $\gamma_{23}$, respectively.

Another effective contribution to the actions of TMG and the GMMG is related to the Chern-Simons terms.
By modification of the bulk description by CS-term, the minimal EWCS can get a contribution from the CS-term. In the flat-TMG/GCFT case, we are encountered with two central charges and the $c_L$, in addition to $c_M$, becomes non-zero. The non-zero $c_L$ corresponds to the fact that the primary twist operators involved in determining the n-point correlation functions have non-zero spins \cite{sp01,sp02}. These operators correspond to spinning particles in the bulk space. The spinning particles move in a straight line in the Minkowski spacetime (based on the proof made in \cite{sp01}). These straight lines are the same extremized curves that are considered in calculating the entanglement wedge. The effect of the Chern-Simons term is that a normal frame is mounted at any point on the bulk \cite{sp01,sp02}. This frame is expressed in terms of the coordinates, and timelike vector $v$ and spacelike vector $\tilde{v}$ that are normal to the trajectory of the particle in the bulk. Thus, the action of a moving particle is not only related to the curve length that connects the endpoints of the trajectory $X^{\mu}$, but also to another factor, which is related to this frame changes. Therefore, the action also depends on the product of normal vectors $v_i.v_f$, which is obtained by the parallel transporting $v_f$ to $v_i$ along the $X^{\mu}$.
This shift in the action is presented as a boost, which is reflected in the following relation \cite{sp02}
\begin{equation}\label{spin01}
  \Delta\eta(v_i,v_f)=\int_{C}ds (\tilde{v}.\nabla v)=\cosh^{-1}(-v_i.v_f).
\end{equation}

Since the minimal EWCS in the first case with $c_L=0$ just depends on the minimal length of the curve homologous to the intervals, when $c_L$ is non-zero, the Chern-Simons boosts are added to the particle action. The CS-contribution to the extremal EWCS can be found as follows
\begin{equation}\label{et01}
  E_W^{CS}=\frac{c_L}{12}\Delta\eta^{extr}(v_b,v'_b),
\end{equation}
where $\Delta\eta^{extr}$ is the extremized boost that is needed to take the parallel transport $v'_b$ and $v_b$ along the bulk entanglement wedge.

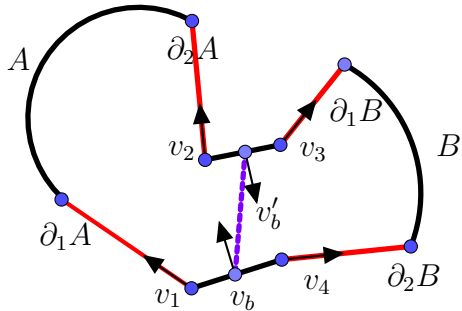
\begin{figure}[!h]
\centering
\captionsetup{width=.8\linewidth}
\definecolor{xfqqff}{rgb}{0.4980392156862745,0,1}
\definecolor{ffqqqq}{rgb}{1,0,0}
\definecolor{xdxdff}{rgb}{0.49019607843137253,0.49019607843137253,1}
\definecolor{ududff}{rgb}{0.30196078431372547,0.30196078431372547,1}
\begin{tikzpicture}[line cap=round,line join=round,>=triangle 45,x=1cm,y=1cm]
\clip(-3.6,-1) rectangle (3,4);
\draw [shift={(0.2818181818181815,1.19)},line width=2pt]  plot[domain=-0.37185607384858166:1.0743735733900155,variable=\t]({1*1.951562205342765*cos(\t r)+0*1.951562205342765*sin(\t r)},{0*1.951562205342765*cos(\t r)+1*1.951562205342765*sin(\t r)});
\draw [shift={(-1.5181818181818179,2.19)},line width=2pt]  plot[domain=1.0684939486969438:3.9551965009733876,variable=\t]({1*1.4728395018966276*cos(\t r)+0*1.4728395018966276*sin(\t r)},{0*1.4728395018966276*cos(\t r)+1*1.4728395018966276*sin(\t r)});
\draw [line width=1.6pt,color=ffqqqq] (-2.542975206611571,1.1057024793388348)-- (-0.8157024793388428,-0.07611570247934825);
\draw [line width=2pt,color=ffqqqq] (-0.8090909090909091,3.4809090909090905)-- (-0.6338842975206608,1.6329752066115628);
\draw [line width=2pt,color=ffqqqq] (1.2199840703542635,2.901268478909613)-- (0.36611570247933994,1.832975206611563);
\draw [line width=2pt,color=ffqqqq] (2.1,0.4809090909090906)-- (0.3842975206611581,0.30570247933883365);
\draw [line width=2pt] (-0.8157024793388428,-0.07611570247934825)-- (0.3842975206611581,0.30570247933883365);
\draw [line width=2pt] (-0.6338842975206608,1.6329752066115628)-- (0.36611570247933994,1.832975206611563);
\draw [line width=2pt,dotted,color=xfqqff] (-0.10241576605212899,1.7392689129052692)-- (-0.23558990898236942,0.10846556990680227);
\draw (-2.979338842975209,0.932975206611562) node[anchor=north west] {$\partial_1A$};
\draw (-1.2884297520661168,3.4057024793388373) node[anchor=north west] {$\partial_2A$};
\draw (0.8752066115702484,2.5875206611570185) node[anchor=north west] {$\partial_1B$};
\draw (1.6388429752066125,0.3875206611570159) node[anchor=north west] {$\partial_2B$};
\draw (2.2933884297520675,2.114793388429745) node[anchor=north west] {$B$};
\draw (-3.4157024793388455,3.260247933884292) node[anchor=north west] {$A$};
\draw (-1.4520661157024806,0.060247933884288264) node[anchor=north west] {$v_1$};
\draw [->,line width=0.8pt] (-0.6338842975206608,1.6329752066115628) -- (-0.7055969487883248,2.3893407171516436);
\draw [->,line width=0.8pt] (0.36611570247933994,1.832975206611563) -- (0.8491998309067388,2.4373722586271414);
\draw [->,line width=0.8pt] (0.3842975206611581,0.30570247933883365) -- (1.2184606940788405,0.3908867725780234);
\draw (-1.2702479338842987,2.023884297520654) node[anchor=north west] {$v_2$};
\draw (0.47520661157024807,2.005702479338836) node[anchor=north west] {$v_3$};
\draw (0.5297520661157027,0.24206611570247027) node[anchor=north west] {$v_4$};
\draw (-0.45206611570247984,0.023884297520651864) node[anchor=north west] {$v_b$};
\draw (-0.10661157024793415,1.2784297520661079) node[anchor=north west] {$v'_b$};
\draw [->,line width=0.8pt] (-0.23558990898236942,0.10846556990680227) -- (-0.45206611570247984,0.8238842975206527);
\draw [->,line width=0.8pt] (-0.10241576605212899,1.7392689129052692) -- (0.05702479338842961,1.042066115702471);
\draw [->,line width=0.8pt] (-0.8157024793388428,-0.07611570247934825) -- (-1.4652222048963053,0.36829253079681046);
\begin{scriptsize}
\draw [fill=ududff] (2.1,0.4809090909090906) circle (2.5pt);
\draw [fill=ududff] (-0.8090909090909091,3.4809090909090905) circle (2.5pt);
\draw [fill=ududff] (-2.542975206611571,1.1057024793388348) circle (2.5pt);
\draw [fill=xdxdff] (1.2199840703542635,2.901268478909613) circle (2.5pt);
\draw [fill=ududff] (-0.8157024793388428,-0.07611570247934825) circle (2.5pt);
\draw [fill=ududff] (-0.6338842975206608,1.6329752066115628) circle (2.5pt);
\draw [fill=ududff] (0.36611570247933994,1.832975206611563) circle (2.5pt);
\draw [fill=ududff] (0.3842975206611581,0.30570247933883365) circle (2.5pt);
\draw [fill=xdxdff] (-0.10241576605212899,1.7392689129052692) circle (2.5pt);
\draw [fill=xdxdff] (-0.23558990898236942,0.10846556990680227) circle (2.5pt);
\end{scriptsize}
\end{tikzpicture}

\caption{The entanglement wedge construction for two disjoint intervals in a $GCFT_2$ dual to flat TMG. In this figure, the CS contribution to the $X^{\mu}$ bulk trajectory of the spinning particle has been depicted by the timelike vectors that are the ingredients of the normal frames in each points of the bulk }
\label{figEW03}
\end{figure}

Given the effect that Chern-Simons term has, in the construction of the entanglement wedge in the $GCFT_2$ dual to the flat TMG, there is a new ingredient compared to the Einstein gravity case. The ingredients are timelike normal vectors at each bulk point that have been depicted in Fig.\ref{figEW03}.

The CS-contribution in the minimal EWCS can be obtained by extremizing the boost $\Delta \eta$ that is needed to drag the normal frame through the timelike vectors $v_b$ and $v'_b$. Since the action structure of the two models TMG and GMMG is similar in terms of Chern-Simons terms, we find that a similar CS-term contribution naturally appears in the minimal EWCS in the flat $GMMG/GCFT$ duality as well.

In this paper, instead of describing the bulk geometry with the TMG, we want to describe it with the GMMG. The GMMG as an extension of the pure Einstein gravity has three important differences \cite{GMMG01}: The CS deformation term, the higher derivative term and the term  with a negative cosmological constant. The GMMG model and the TMG model
are similar in some respects in that they include the Chern-Simons and the Einstein gravity
parts, but the former includes the higher derivative gravity terms.
Despite the differences, we find that in the GMMG case, the extremal EWCS must consist of two parts, one similar to (\ref{el01}) and the other similar to (\ref{et01}). Although the first term in our expression involves more than pure Einstein gravity due to the higher-derivative terms in GMMG, we present an argument in Appendix \ref{app} that supports the validity of our overall analytic derivation. Meanwhile, the CS-part of the action in the GMMG is not the same in all respects as the CS-term in TMG, and there are some differences. The expression we present in this paper has been analytically derived, but in the rest of this article, we provide evidences that it is true. The extremal EWCS for the flat GMMG that is dual to GCFT with two bipartite intervals  is consistent with our previous calculations of the holographic negativity in the flat holography scenario in \cite{GMMG02} and this can be considered as a strong evidence for the correctness of our result.

For two disjoint intervals $A$ and $B$ in the $GCFT_2$ vacuum that is dual to three dimensional Minkowski spacetime (\ref{Mink01}), the minimal curve length between $p_b$ and $p'_b$ can be considered in the pure Einsein gravity case as follows \cite{Seng03}
\begin{equation}\label{Le01}
  L^{extr}(p_b,p'_b)=\bigg|\frac{\mathcal{X}}{\sqrt{\mathcal{T}}(1-\mathcal{T})}   \bigg|
\end{equation}
which needs to be proven independently in the flat GMMG, because there is the higher derivative term in the GMMG than the Einstein gravity case. $\mathcal{T}$ and $\mathcal{X}/\mathcal{T}$ are the cross-ratios that can be defined as follows
\begin{eqnarray}\label{CrR01}
  \mathcal{T} &=& \frac{t_{12}t_{34}}{t_{13}t_{24}},\nonumber \\
  \frac{\mathcal{X}}{\mathcal{T}} &=& \frac{x_{12}}{t_{12}}+\frac{x_{34}}{t_{34}}-\frac{x_{13}}{t_{13}}-\frac{x_{24}}{t_{24}}.
\end{eqnarray}

In the limit $\mathcal{T}\to1$ which is characteristic of the t-channel computations in the monodromy analysis, we can find the minimal EWCS for the bulk dual of the GCFT with mixed bipartite disjoint intervals in the vacuum as follows
\begin{eqnarray}\label{el02}
  E_{W1} &=& \frac{c_M}{12}L^{extr}(p_b,p'_b) \nonumber\\
  &=& -\frac{1}{4G}(\sigma+\frac{\alpha H}{\mu}+\frac{F}{m^2})L^{extr}(p_b,p'_b), \nonumber\\
    &=& -\frac{1}{4G}(\sigma+\frac{\alpha H}{\mu}+\frac{F}{m^2}) \big(\frac{\mathcal{X}}{\sqrt{\mathcal{T}}(1-\mathcal{T})} \big),\nonumber   \\
    &=& -\frac{1}{4G}(\sigma+\frac{\alpha H}{\mu}+\frac{F}{m^2})\big(\frac{x_{13}}{t_{13}}+\frac{x_{24}}{t_{24}}-\frac{x_{14}}{t_{14}}-\frac{x_{23}}{t_{23}}   \big),
\end{eqnarray}
where we have substituted (\ref{cLM-GMMG}) into (\ref{el01}) for the GMMG model. The minimal boost between $v_b$ and $v'_b$ can preserve its shape as in the flat TMG case as follows \cite{Seng03}
\begin{equation}\label{et02}
  \Delta\eta^{extr}=2\log\frac{1+\sqrt{\mathcal{T}}}{\sqrt{1-\mathcal{T}}},
\end{equation}
where in the t-channel limit ($\mathcal{T}\to1$), the above equation can be chaned as follows
\begin{equation}\label{et03}
  \Delta\eta^{extr}=2\log 2-\log (1-\mathcal{T}).
\end{equation}

Substituting (\ref{et03}) and the $c_L$ central charge (\ref{cLM-GMMG}) into (\ref{et01}), we find the CS-contribution to the minimal EWCS for two disjoint intervals in the vacuum as follows
\begin{equation}\label{et04}
  E_W^{CS} = \frac{1}{4\mu G}\log \frac{t_{13}t_{24}}{t_{14}t_{23}}.
\end{equation}

Using (\ref{el02}) and (\ref{et04}), we find the minimal EWCS in the asymptotically flat space bulk in the GMMG that is dual to the two intervals in the $GCFT_2$ as follows
\begin{eqnarray}\label{Res01}
  E_W &=& E_{W1}+E_W^{CS}\nonumber \\
   &=& -\frac{1}{4G}(\sigma+\frac{\alpha H}{\mu}+\frac{F}{m^2})\big(\frac{x_{13}}{t_{13}}+\frac{x_{24}}{t_{24}}-\frac{x_{14}}{t_{14}}-\frac{x_{23}}{t_{23}}   \big) + \frac{1}{4\mu G}\log \frac{t_{13}t_{24}}{t_{14}t_{23}}.
\end{eqnarray}

Substituting the result (\ref{Res01}) on the minimal EWCS into the proposal (\ref{ENEW01}), the holographic entanglment negativity $\mathcal{E}$ of two intervals at zero temperature $GCFT_2$ that is dual to the flat GMMG, can be found as follows
\begin{equation}\label{HEN02}
  \mathcal{E}=
   -\frac{3}{8G}(\sigma+\frac{\alpha H}{\mu}+\frac{F}{m^2})\big(\frac{x_{13}}{t_{13}}+\frac{x_{24}}{t_{24}}-\frac{x_{14}}{t_{14}}-\frac{x_{23}}{t_{23}}   \big) + \frac{3}{8\mu G}\log \frac{t_{13}t_{24}}{t_{14}t_{23}}.
\end{equation}

This result is the same as the holographic entanglement negativity for two disjoint intervals configurations at zero
temperature holographic $GCFT_2$ in \cite{GMMG02}.

\subsection{EWCS for two disjoint intervals at a finite temperature }\label{sec:3.2}
In this section, the two intervals $A$ and $B$ are considered in a thermal $GCFT_2$ on a cylinder with the temperature that is equal to its inverse circumference. The bulk metric can be given by  three dimensional flat space cosmology as follows \cite{GC04,GC05,GC07}
\begin{equation}\label{FSC01}
  ds^2=Mdu^2-2du dr+r^2d\phi^2,
\end{equation}
where $M$ has the following relation with the temperature of the $GCFT_2$
\begin{equation}\label{M01}
  M=\big(\frac{2\pi}{\beta}  \big)^2.
\end{equation}

If the bulk geometry is considered by the pure Einstein gravity, the minimal curve length between $p_b$ and $p'_b$ in the t-channel limit ($T\to1$) can be found as follows \cite{Seng03}

\begin{eqnarray}\label{elT01}
  L^{extr}(p_b,p'_b)&=&\frac{\pi u_{13}}{\beta}\coth\big(\frac{\pi\phi_{13}}{\beta} \big)
  +\frac{\pi u_{24}}{\beta}\coth\big(\frac{\pi\phi_{24}}{\beta} \big) \nonumber\\
  &-&\frac{\pi u_{14}}{\beta}\coth\big(\frac{\pi\phi_{14}}{\beta} \big)
  -\frac{\pi u_{23}}{\beta}\coth\big(\frac{\pi\phi_{23}}{\beta} \big),
\end{eqnarray}
where it is needed to be proved in the GMMG scenario with the higher derivative deformation in related to the pure Einstein gravity. Using the corresponding central charge (\ref{cLM-GMMG}) in the $GCFT_2$ which is dual to the GMMG in 3d asymptotically flat spacetime and the relation (\ref{el01}), the extremal EWCS in the GMMG case, regardless of the CS-contribution, can be found as follows
\begin{eqnarray}\label{elT02}
  E_{W1} &=& -\frac{1}{4G}(\sigma+\frac{\alpha H}{\mu}+\frac{F}{m^2}) \bigg [\frac{\pi u_{13}}{\beta}\coth\big(\frac{\pi\phi_{13}}{\beta} \big) \nonumber \\
   &+& \frac{\pi u_{24}}{\beta}\coth\big(\frac{\pi\phi_{24}}{\beta} \big)
  -\frac{\pi u_{14}}{\beta}\coth\big(\frac{\pi\phi_{14}}{\beta} \big)
  -\frac{\pi u_{23}}{\beta}\coth\big(\frac{\pi\phi_{23}}{\beta} \big)\bigg ].
\end{eqnarray}

To consider the CS-contribution in the minimal EWCS, the timelike vectors as new ingredients can be changed in the FSC geometry \cite{Seng02,Seng03}. We assume that the minimal boost between $v_b$ and $v'_b$ can have a similar relation to the TMG case as follows \cite{Seng03}
\begin{equation}\label{etT01}
  \Delta\eta^{extr}=\log\bigg[\frac{\sinh\frac{\pi\phi_{13}}{\beta}\sinh\frac{\pi\phi_{24}}{\beta}}
  {\sinh\frac{\pi\phi_{14}}{\beta}\sinh\frac{\pi\phi_{23}}{\beta}}   \bigg]
\end{equation}

Using (\ref{etT01}) and (\ref{cLM-GMMG}) into (\ref{et01}), the CS-contribution in the minimal EWCS for the holographic dual of the $GCFT_2$ at finite temperature can be found as follows
\begin{equation}\label{etT02}
  E_W^{CS}=\frac{1}{4\mu G}\log\bigg[\frac{\sinh\frac{\pi\phi_{13}}{\beta}\sinh\frac{\pi\phi_{24}}{\beta}}
  {\sinh\frac{\pi\phi_{14}}{\beta}\sinh\frac{\pi\phi_{23}}{\beta}}   \bigg].
\end{equation}

Using the minimal EWCS (\ref{elT02}) and its CS-modification (\ref{etT02}) that are in the GMMG scenario, the final form of the extremized EWCS at the finite temperature case can be find as follows

\begin{eqnarray}\label{Res02}
  E_W &=& E_{W1}+E_W^{CS} \nonumber\\
   &=& -\frac{1}{4G}(\sigma+\frac{\alpha H}{\mu}+\frac{F}{m^2}) \bigg [\frac{\pi u_{13}}{\beta}\coth\big(\frac{\pi\phi_{13}}{\beta} \big)+\frac{\pi u_{24}}{\beta}\coth\big(\frac{\pi\phi_{24}}{\beta} \big) \nonumber \\
   &-&\frac{\pi u_{14}}{\beta}\coth\big(\frac{\pi\phi_{14}}{\beta} \big)
  -\frac{\pi u_{23}}{\beta}\coth\big(\frac{\pi\phi_{23}}{\beta} \big)\bigg ]
  +\frac{1}{4\mu G}\log\bigg[\frac{\sinh\frac{\pi\phi_{13}}{\beta}\sinh\frac{\pi\phi_{24}}{\beta}}
  {\sinh\frac{\pi\phi_{14}}{\beta}\sinh\frac{\pi\phi_{23}}{\beta}}   \bigg]
\end{eqnarray}

Substituting the result (\ref{Res02}) on the minimal EWCS into the proposal (\ref{ENEW01}), the holographic entanglment negativity $\mathcal{E}$ of two intervals at finite temperature $GCFT_2$ that is dual to the flat GMMG, can be found as follows
\begin{eqnarray}\label{HEN03}
  \mathcal{E} &=& -\frac{3}{8G}(\sigma+\frac{\alpha H}{\mu}+\frac{F}{m^2}) \bigg [\frac{\pi u_{13}}{\beta}\coth\big(\frac{\pi\phi_{13}}{\beta} \big)+\frac{\pi u_{24}}{\beta}\coth\big(\frac{\pi\phi_{24}}{\beta} \big) \nonumber \\
   &-&\frac{\pi u_{14}}{\beta}\coth\big(\frac{\pi\phi_{14}}{\beta} \big)
  -\frac{\pi u_{23}}{\beta}\coth\big(\frac{\pi\phi_{23}}{\beta} \big)\bigg ]
  +\frac{3}{8\mu G}\log\bigg[\frac{\sinh\frac{\pi\phi_{13}}{\beta}\sinh\frac{\pi\phi_{24}}{\beta}}
  {\sinh\frac{\pi\phi_{14}}{\beta}\sinh\frac{\pi\phi_{23}}{\beta}}   \bigg]
\end{eqnarray}

This result is the same as the holographic entanglement negativity for two disjoint intervals configurations at finite
temperature holographic $GCFT_2$ in \cite{GMMG02}.

\section{Conclusion  }\label{sec:4}
In this paper, we considered the minimal entanglement wedge cross section (EWCS) on a flat holography scenario.
We have derived an analytic expression for the minimal EWCS for the bipartite mixed states in $GCFT_2$, and provided strong evidence for its correctness by comparing with previous results on holographic negativity.
Based on the argument presented in the appendix \ref{app} and our analytic derivation, we conclude that the
minimal EWCS of the GMMG in the context of the asymptotically flat geometry can have
a similar structure with the minimal EWCS in the flat TMG. This argument is based on a comparison of the holographic entanglement entropy in the two models and can be used to support the validity of the result we have derived.
The GMMG model and the TMG model are
similar in some respects in that they include the Chern-Simons and the Einstein gravity
parts, but the former includes the higher derivative gravity terms. Despite the differences,
we find that in the GMMG case, the extremal EWCS must consist of two parts, one similar
to (\ref{el01}) and the other similar to (\ref{et01}).
The first part is related to the minimal EWCS in a pure Einstein gravity case and the second part is related to the minimal EWCS contribution from the Chern-Simons (CS-) term of the model.
We conclude that the minimal EWCS approach offers a viable and consistent method to compute the holographic entanglement negativity in flat space holography.
To support this claim, we argue that the minimal EWCS and the holographic entanglement negativity in flat holography such as the $AdS_3/CFT_2$ could satisfy the relation (\ref{ENEW01}).
The results obtained for the holographic entanglement negativity in the two cases we examined, in relations (\ref{HEN02}) and (\ref{HEN03}), are the same as our previous results, which we obtained previously using an independent approach in \cite{GMMG02}. This correspondence supports the validity of our analytic derivation.

This successful application of the EWCS framework in GMMG opens up avenues for exploring other information-theoretic quantities, such as reflected entropy or complexity, in the context of flat holography. Furthermore, extending this analysis to include higher-spin theories or different boundary configurations remains a compelling direction for future research.

\begin{appendices}
\section{ Comparison of the entanglement entropies of GMMG and TMG }\label{app}
Both the holographic entanglement entropy and the minimal EWCS are related to the geometric calculations inside the bulk. From this point of view, if we can provide an argument based on the holographic entanglement entropy, we argue that the results can be used for the minimal EWCS. As mentioned in the section \ref{sec:3}, the GMMG and TMG models are similar as they both include Chern-Simons and Einstein-Hilbert terms; however, GMMG also contains higher-derivative gravity terms. Therefore, comparing the entanglement entropy of these models in the context of the flat holography can have good results. We perform the computation in this Appendix by considering the pure state of a single interval in a $GCFT_2$ at the boundary is dual to a gravity on the asymptotically flat background. We examine the bulk geometry once by the TMG and again by the GMMG.

The holographic entanglement entropy for the TMG in an asymptotically flat background which is dual to the $GCFT_2$ at the boundary with a single interval can be found in the literature as follows \cite{GC05}
\begin{equation}\label{EEA01}
  S_{EE}=\frac{c'_M}{12}\bigg( u_{12}\cot\big( \frac{\phi_{12}}{2}  \big)     \bigg)+\frac{c'_L}{12}\bigg(2\log\big(\frac{2}{\epsilon}\sin\frac{\phi_{12}}{2}  \big)   \bigg),
\end{equation}
where $c'_L$ and $c'_M$ are the central charges of the $GCFT_2$ dual to TMG in an asymptotically flat background. By comparing the calculations performed in \cite{Seng03}, it can be seen that
\begin{eqnarray}\label{LEtA01}
  \Delta\eta^{extr} &=& 2\log\big(\frac{2}{\epsilon}\sin\frac{\phi_{12}}{2}  \big),\\
  L^{extr} &=& u_{12}\cot\big( \frac{\phi_{12}}{2}  \big), \label{LEtA02}
\end{eqnarray}

that $L^{extr}$ and $\Delta\eta^{extr}$ are the minimal length of the curve homologous to the entangling interval and the extremized boost respectively computed for a single interval configuration at the boundary with $(u_1,\phi_1)$ and $(u_2,\phi_2)$ endpoints. By substituting the $L^{extr}$ and $\Delta\eta^{extr}$ into (\ref{EEA01}), we can change the form of the TMG holographic entanglement entropy as follows
\begin{equation}\label{EEA02}
  S_{EE}=\frac{c'_M}{12}L^{extr}+\frac{c'_L}{12}\Delta\eta^{extr},
\end{equation}
where this result is obtained if we have a GCFT with zero temperature at the boundary and the global Minkowski spacetime in the bulk. We have considered such a configuration for the GMMG model and the results are presented in \cite{HEESK}. In this paper, the holographic entanglement entropy is computed for the asymptotically flat GMMG  which is dual to a field theory whose the symmetry generators in the vacuum satisfy the $GCFT_2$ generators algebra. The GCFT is considered at a zero temperature configuration with a single interval as an entangling surface. The entanglement entropy for the mentioned configuration of the $GMMG/GCFT$ duality is computed as follows \cite{HEESK}
\begin{equation}\label{EEA03}
  S_{EE}=-\frac{1}{4G}\big(\sigma+\frac{\alpha H}{\mu}+\frac{F}{m^2}   \big)\bigg(u_{12}\cot\big( \frac{\phi_{12}}{2}  \big)\bigg)+\frac{1}{4\mu G}\bigg(2\log\big(\frac{2}{\epsilon}\sin\frac{\phi_{12}}{2}  \big)  \bigg).
\end{equation}

By substituting the relation (\ref{LEtA01}) and (\ref{LEtA02}) into the holographic entanglement entropy (\ref{EEA03}), we have the following relation
\begin{equation}\label{EEA04}
  S_{EE}=\frac{c_M}{12}L^{extr}+\frac{c_L}{12}\Delta\eta^{extr},
\end{equation}
where we have used the central charge relations (\ref{cLM-GMMG}) into (\ref{EEA03}). Comparing the relations (\ref{EEA02}) and (\ref{EEA04}), we find that the holographic entanglement entropy for the flat-TMG and the flat-GMMG for a single interval configuration has a similar structure. If we consider $L^{extr}$ and $\Delta\eta^{extr}$ as the result of geometric calculations, we assume that these calculations reach the same result in both dualities. The effect of the differences between GMMG and TMG is reflected only in the central charges in the result of the holographic entanglement entropy.

The computation of $L^{extr}$ and $\Delta\eta^{extr}$ in the cases where the gravity is dual with the GCFT with a single interval configuration at the boundary is no different in holographic entanglement entropy case and the minimal EWCS case. We conclude that in the mixed bipartite configurations the structure remains the same and we can use the results obtained in TMG to reliably derive the corresponding results for GMMG.

\end{appendices}

\section*{Acknowledgments}
M.K. would like to dedicate this work to the memory of his esteemed professor and mentor, Mohammad Reza Setare, who sadly passed away before the completion of this manuscript. His guidance, wisdom, and invaluable contributions were fundamental to this research.
$^\dagger$Deceased.

\end{document}